\documentclass[a4paper,11pt]{article}
\pdfoutput=1 

\usepackage{jheppub} 

\usepackage[T1]{fontenc} 

\usepackage{dsfont}

\title{Correlations and the ridge in the Color Glass Condensate beyond the glasma graph approximation}

\author[a]{Tolga Altinoluk,}
\author[b]{N\'{e}stor Armesto}
\author[b]{and Douglas E. Wertepny}

\affiliation[a]{National Centre for Nuclear Research, 00-681 Warsaw, Poland}
\affiliation[b]{Departamento de F\'{i}sica de Part\'{i}culas and IGFAE, \\Universidade de Santiago de Compostela, 15782 Santiago de Compostela, Galicia-Spain}

\emailAdd{tolga.altinoluk@ncbj.gov.pl}
\emailAdd{nestor.armesto@usc.es}
\emailAdd{douglas.wertepny@usc.es}

\abstract{We consider two-gluon production in dilute-dense collisions within the Color Glass Condensate framework, applicable to both proton-nucleus and heavy-light ion collisions. We go beyond the glasma graph approximation which is valid in the dilute-dilute limit and show the correspondence between the glasma graphs and the $k_T$-factorized approach that we use in our calculation. We then identify the classical uncorrelated, and the Hanbury-Brown-Twiss (HBT) and Bose enhancement correlated contributions, with the Bose enhancement contribution being suppressed by the number of degrees of freedom with respect to the uncorrelated piece. We show that both the HBT and the Bose enhancement pieces survive the inclusion of higher order contributions in density and that they stem from the quadrupole piece of the two-gluon inclusive cross section. Finally, we illustrate the results using a toy model that allows a simple numerical implementation.}

\begin{document} 
\maketitle
\flushbottom


\section{Introduction}
\label{sec:intro}

The ridge phenomenon - the existence of long range pseudorapidity two-particle correlations peaking for identical or opposite azimuthal directions - is one of the main findings of the Relativistic Heavy Ion Collider (RHIC) at BNL and the Large Hadron Collider (LHC) at CERN concerning Quantum Chromodynamics (QCD). First discovered in AuAu collisions at RHIC \cite{Alver:2009id,Abelev:2009af}, it was later found in pp collisions at the LHC \cite{Khachatryan:2010gv,Khachatryan:2015lva,Aad:2015gqa} and subsequently observed in all collisional systems, including the small ones (pp, pPb, dAu, $^3$HeAu) for high multiplicity events \cite{CMS:2012qk,Abelev:2012ola,Aad:2012gla,Aaij:2015qcq,Khachatryan:2016ibd,Adare:2014keg,Adamczyk:2015xjc,Adare:2015ctn}. More recently, sizeable azimuthal anisotropies have also been observed in small systems for events with multiplicities much closer to average \cite{Khachatryan:2016txc,Aaboud:2016yar,Aaboud:2017acw,Aaboud:2017blb}.

The standard explanation of the ridge in heavy-ion collisions comes through the coupling of an initial long range pseudorapidity correlation to an expanding medium. This medium is accurately described by viscous relativistic hydrodynamics \cite{Bozek:2012gr,Shuryak:2013ke,Bzdak:2013zma,Werner:2010ss,Gavin:2008ev}. Nevertheless, the validity of the assumptions underlying the hydrodynamical explanation becomes tenuous for small systems where isotropization and smallness of the mean free path are difficult to justify. One alternative currently under exploration is that hydrodynamics seems to be applicable for out-of-equilibrium systems. This has been argued in both weak and strong coupling approaches \cite{Chesler:2009cy,Heller:2011ju,Kurkela:2015qoa}. The ridge phenomenon is thus a key observable for our understanding of the emergence of a macroscopic description in hadronic and nuclear collisions from the underlying QCD microscopic dynamics \cite{Romatschke:2016hle}.

On the other hand, explanations alternative to hydrodynamics exist that may shed light on the emergence problem and, in any case, should be used to provide the initial conditions for and characterise the dynamics prior to hydrodynamic evolution. In this work we focus on those given by the weak coupling but non-perturbative realisation of QCD at high energies provided by the Color Glass Condensate (CGC) effective field theory \cite{Gelis:2010nm}, but other approaches exist based on different non-perturbative ideas, see e.g. \cite{Hwa:2008um,Bjorken:2013boa,Shuryak:2013sra,Andres:2014bia}.

In the CGC framework, the ridge was phenomenologically addressed through an approximation valid when both projectile and target are dilute, called the ``glasma graphs" \cite{Armesto:2006bv,Dumitru:2008wn,Dumitru:2010iy}, see \cite{Kovchegov:2012nd,Kovchegov:2013ewa} for an analogous calculation in slightly different language. Such an approximation for gluon-gluon correlations, assuming that it can be translated to the final particles, was used to successfully describe the measurements in pp \cite{Dusling:2012iga,Dusling:2012cg}, and was later extended to pA collisions in various phenomenological ways \cite{Dusling:2012wy,Dusling:2013qoz,Dusling:2017dqg,Dusling:2017aot}. The dilute-dilute results have been argued recently to stem from coherence in parton radiation \cite{Blok:2017pui} without any reliance on the CGC formalism. The extension to high densities has been explored recently to quantify the validity of the glasma graph approach and to analyze the odd Fourier harmonics in the azimuthal particle correlations \cite{Skokov:2014tka,Schenke:2015aqa,Lappi:2015vta,McLerran:2016snu,Kovner:2016jfp,Kovchegov:2018jun}. It has also been extended to partons other than gluons \cite{Altinoluk:2016vax,Kovner:2017gab,Martinez:2018ygo} and to the forward region linking with the multiple parton scattering language \cite{Kovner:2017vro,Kovner:2017ssr,Kovner:2018vec}. On the other hand, other ideas within the CGC framework exist, considering the existence of domains of oriented chromoelectric fields in the hadron or nucleus \cite{Kovner:2010xk,Kovner:2011pe,Kovner:2012jm,Dumitru:2014vka}, or justifying the azimuthal correlations through the density profile of the hadron \cite{Levin:2011fb}.

In \cite{Altinoluk:2015uaa}, the origin of the ridge azimuthal correlations in the glasma graph approach was identified to come from the Bose enhancement of gluons in the wave function of the incoming hadrons. Similar calculations also showed the existence of Hanbury-Brown-Twiss (HBT) correlations of the produced particles \cite{Kovchegov:2012nd,Kovchegov:2013ewa,Altinoluk:2015eka}. The aim of the present work is to establish whether the Bose enhancement contribution found in the glasma graph approach survives the density corrections that appear in the dilute-dense situation and, if so, to identify which contributions from the color ensembles in the target are dominant. We anticipate that our answer is positive and that, somewhat unexpectedly (although previously claimed in e.g. \cite{Kovner:2017ssr}), we find that the contribution to the Bose enhancement terms (also to the HBT ones) comes from the quadrupole distribution of Wilson lines in the target. The contribution from the target average of two dipoles turns out to be suppressed by a relative factor of $\frac{1}{N_c^2-1}$.

The plan of the paper is as follows: In Section \ref{section2} we present the setup and describe the previous results on Bose-enhanced contributions to the ridge. Section \ref{section3} contains our main results. In Section \ref{toymodel} we illustrate them using a toy model. Finally, in Section \ref{conclusions} we summarise and present our conclusions.


\section{Setup and previous results}
\label{section2}

\subsection{Bose Enhancement and the ridge}

\begin{figure}[tbp]
\centering 
\includegraphics[width=.3\textwidth]{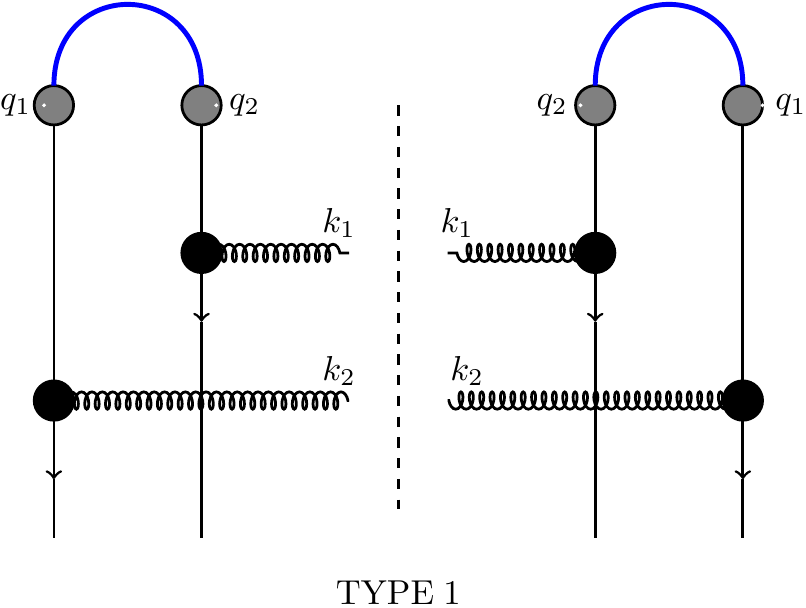}
\hfill
\includegraphics[width=.3\textwidth]{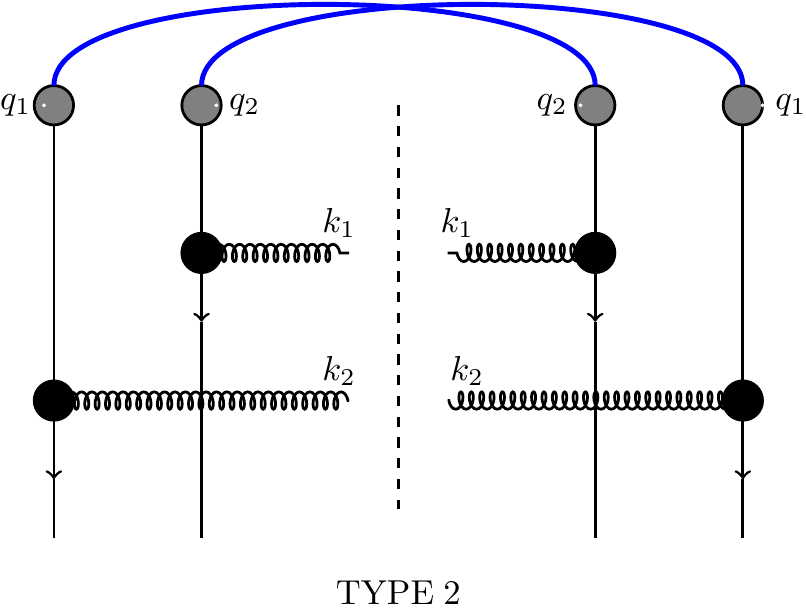}
\hfill
\includegraphics[width=.3\textwidth]{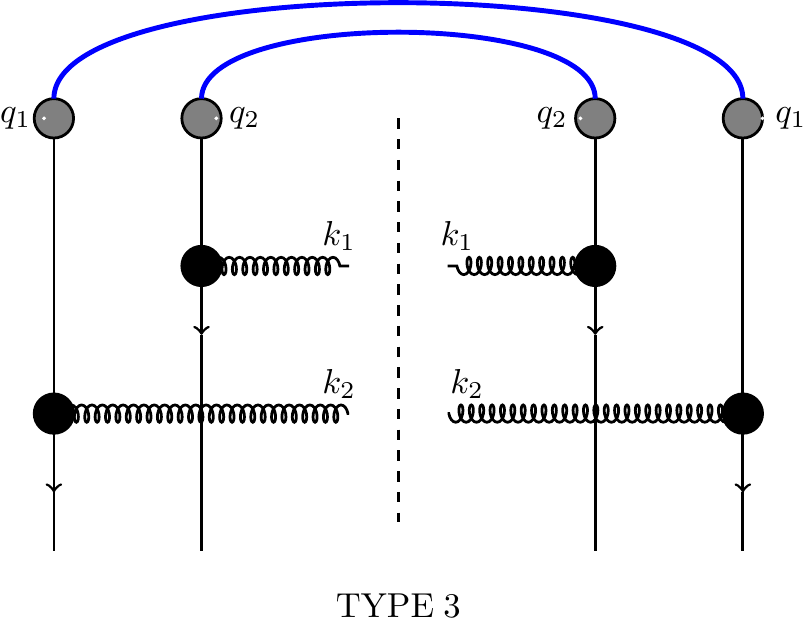}
\caption{\label{fig:1} Diagrams corresponding to the glasma graph calculation of two-particle correlations. See the text for explanations.}
\end{figure}

As mentioned in the Introduction, in the Color Glass Condensate particle correlations have been studied for phenomenological purposes using the glasma graph approach, which has produced successful comparisons with experimental data. In this approach, it was shown in \cite{Kovchegov:2012nd,Kovchegov:2013ewa,Altinoluk:2015uaa,Altinoluk:2015eka} that the ridge receives two contributions:

\begin{itemize}
\item One contribution comes from the Bose enhancement of the gluons in the projectile wave function, that results in a form of the normalised two-particle correlation of gluons in the projectile wave function
\begin{equation}
\label{eq1}
D(q_1,q_2)\propto 1+ \frac{1}{S_\perp (N_c^2-1)} \left[ \delta^{(2)}(q_1-q_2)+\delta^{(2)}(q_1+q_2)\right],
\end{equation}
with $q_1,q_2$ the transverse momentum of the gluons, see Fig. \ref{fig:1}. In this equation, 1 stands for the classical, uncorrelated term, and the $\delta$-functions come from the Bose statistics of the partons in the adjoint (real) representation, the gluons. Bose enhancement can be clearly seen by the fact that these latter  are suppressed by the number of degrees of freedom: transverse area $S_\perp$ times number of gluons. The wave function of the projectile in the CGC is boost invariant up to rapidities $y\sim 1/\alpha_s$, large in the weak coupling regime.
This is the correlation in the projectile wave function; in order to get the corresponding cross section, one convolutes with the probability density for scattering of the two gluons in the target, which amounts to multipole distributions that would smear the correlations by momenta of order the saturation scale of the target.
\item Another contribution comes from the HBT correlations in the final state (with gluon momenta $k_1,k_2$), so after rescattering with the target. These correlations are sensitive to the size of the projectile, therefore they appear at smaller distance (in $|k_1\pm k_2|$) from the peak than the previous ones, and are enhanced with respect to them by the number of particle sources ($S_\perp$ times  the saturation scale of the target squared).
\end{itemize}

In Fig. \ref{fig:1} we show the relevant graphs for the computation of correlations in the CGC. Two gluons are emitted from two different color sources (grey blobs) from the projectile, with momentum $q_1$ and $q_2$. The curved blue lines indicate the color contractions of the projectile sources. The diagram on the right, type 3, does not produce any correlations (apart from those that may stem from geometry, see below).  Note that we have not yet performed the rescattering with the target. In the glasma graph approach, such rescattering is done at the lowest order in the target density (two gluon exchange leading to dilute-dilute scattering), with the final state gluons coming from Lipatov vertices (black blobs). Both Bose enhancement and HBT correlations result from the two leftmost diagrams, type 1 and 2.

The aim of this paper is the extension of these calculations to the dilute-dense situation, i.e., beyond the lowest order in target density. For that we will use the language in \cite{Kovchegov:2012nd,Kovchegov:2013ewa}. It was proved there that, after several manipulations, a $k_T$-factorized form can be written for two-particle correlations. In this formalism, analogous results were obtained with geometrical (classical), HBT and Bose contributions. We present the $k_T$-factorized formalism in the next section and its correspondence to the glasma graph approach that we have just discussed.


\subsection{Two-particle correlations in the $k_T$-factorized form}

\begin{figure}[tbp]
\centering
\includegraphics[width=.3\textwidth]{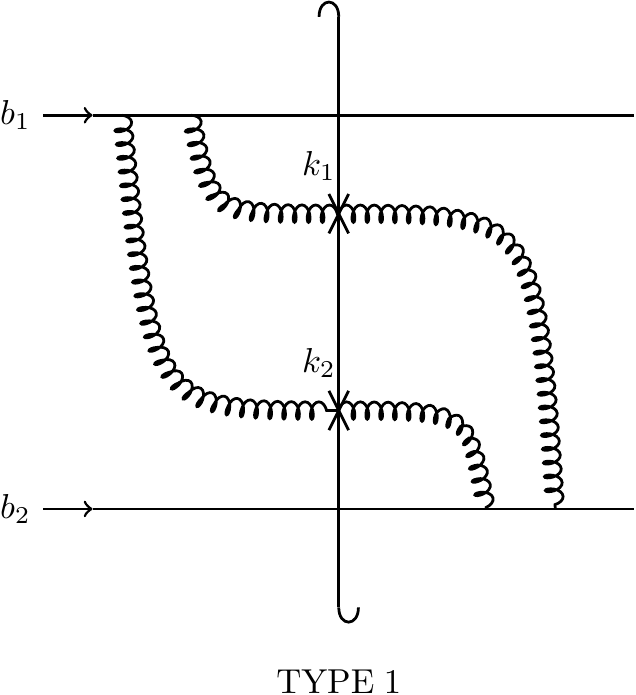}
\hfill
\includegraphics[width=.3\textwidth]{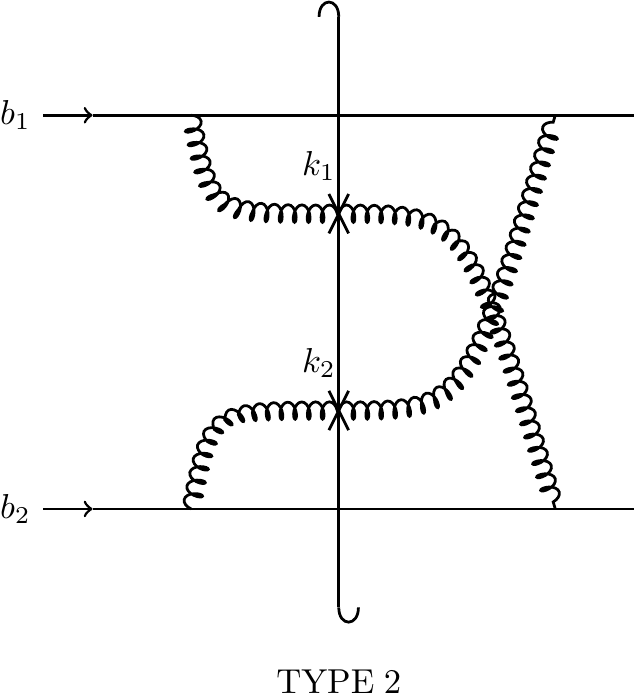}
\hfill
\includegraphics[width=.3\textwidth]{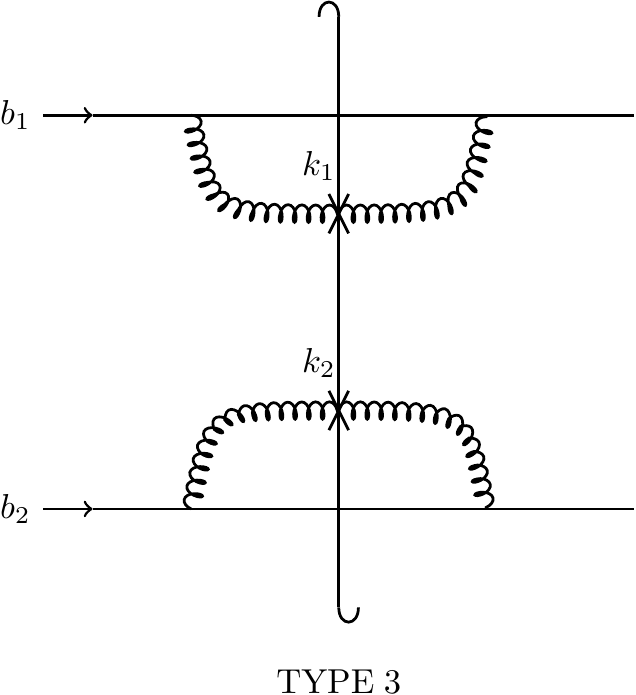}
\caption{\label{fig:2} Diagrams corresponding to the $k_T$-factorized calculation of two-particle correlations. See the text for explanations.}
\end{figure}

The two-gluon correlation function in the dilute-dense (or heavy-light) limit, where  the projectile is defined by two color sources as in the previous Subsection and all rescatterings in the target are taken into account, was originally derived in \cite{Kovchegov:2012nd,Kovchegov:2013ewa} in a $k_T$-factorized form.
Three distinct processes contributed, see Fig. \ref{fig:2}:
\begin{itemize}
\item The ones exemplified on the right, type 3, where each projectile source emitted the same gluon in the amplitude and  in the complex conjugate amplitude. These diagrams, referred to as squared diagrams, correspond exactly to the type 3 ones in Fig. \ref{fig:1} and, at large $N_c$, do not provide any correlations apart from trivial, geometrical ones. 
\item The type 2 ones in which each projectile source is attached to one gluon in the amplitude and to the other gluon in the complex conjugate amplitude. These are part of the connected diagrams and correspond to type 2 diagrams in Fig. \ref{fig:1}.
\item The type 1 ones where one projectile source emitted two gluons in the amplitude that are attached to the other projectile source in the complex conjugate amplitude. They provide the remaining part of the connected diagrams and correspond to type 1 diagrams in Fig. \ref{fig:1}.
\end{itemize}
Diagrams of type 3 contain the double trace of Wilson lines while type 1 and 2 contain the quadrupole contribution.

One should note that, since we have two indistinguishable gluons, this process is invariant under interchange of $\vec k_1$ and $\vec k_2$.
The separated diagrams' contribution is explicitly invariant when $\vec k_2 \rightarrow - \vec k_2$ and diagrams of type 1 become diagrams of type 2 under this interchange. Thus, the total expression is invariant under the interchange of $\vec k_1$ and $\vec k_2$ and  $\vec k_2 \rightarrow - \vec k_2$. This means the two-gluon production cross section is explicitly even and will generate a symmetric, in azimuthal angle, correlation function, a fact that in the previous Subsection resulted from the reality of the gluon color representation. This  can be seen explicitly in the full results presented in the original work and in the equations presented later in this one.

While the form presented in the original work \cite{Kovchegov:2012nd} contains all the information required to see if the Bose correlation terms survive the inclusion of higher order rescatterings in the target, it is rather difficult to extract the terms that correspond to the various correlations.
In a follow up paper \cite{Kovchegov:2013ewa}, a $k_T$-factorized form was derived for the two-gluon production cross section.
Here we will explicitly see the way in which the gluon distribution functions associated with the projectile and target nuclei conspire to produce the correlations.
This form ends up being much easier to manipulate in order to isolate the Bose-enhanced contribution.

Before delving into the equation itself, it is first necessary to define the distribution functions that compose the correlation function.
The $k_T$-factorized form of the single inclusive gluon production cross section involves the usual unintegrated gluon distributions,
\begin{align}
\label{eq:unint_wave} 
\phi_{A_1} ({\vec q}; y) =
  \frac{C_F}{\alpha_s ( 2 \pi)^3} \int d^2 b \, d^2 r \; 
  e^{-i {\vec q} \cdot {\vec r}} \; \nabla_{{\vec r}}^2 \; n_G ({\vec b} + {\vec r}, {\vec b}; y)
\end{align}
for the projectile nucleus and 
\begin{align}
\label{eq:unint_target} 
\phi_{A_2} ({\vec q}; y) =
  \frac{C_F}{\alpha_s ( 2 \pi)^3} \int d^2 b \, d^2 r \; 
  e^{-i {\vec q} \cdot {\vec r}} \; \nabla_{{\vec r}}^2 \; N ({\vec b} + {\vec r}, {\vec b}; y)
\end{align}
for the target.

Note that our wave function is rapidity independent and that in the semiclassical calculation that we are presenting, no quantum evolution is allowed neither between the two gluons not between them and any of the projectile or target hadrons or nuclei. Therefore $y$ is simply a label for the rapidity distance between projectile and target, i.e., for the collision energy, that would correspond to the rapidity at which the target multipole distributions are evaluated and, thus, it is not well determined without computing the quantum evolution.

Here $n_G ({\vec b} + {\vec r}, {\vec b}; y)$ is the distribution associated with the proton (or light nucleus or, in general, dilute projectile) which in our semiclassical calculation is \footnote{In the toy model that we will present later, the saturation scale does not depend on impact parameter $\vec b$.}
\begin{equation}
\label{project_dis}
n_G ({\vec b} + {\vec r}, {\vec b}; y = 0) = \frac{1}{4} Q_{s,1}^2(\vec b) \, r^2 \ln{ \left( \frac{1}{r \, \Lambda} \right) },
\end{equation}
where $\Lambda$ is an infrared (IR) cutoff and $y$ is the rapidity.
Note that usually this is notated as $n_G (\vec x, \vec y; y)$ where the $\vec r = \vec x - \vec y$ and the saturation scale, $Q_{s,1}^2(\vec b)$, is evaluated at $\frac{1}{2}\left( \vec x + \vec y \right)$.
In our case we assume that the saturation scale is slowly varying such that $Q_{s,1}^2(\vec b) \approx Q_{s,1}^2(\vec b +\vec r)$ so the difference between these two notations is negligible. $N ({\vec b} + {\vec r}, {\vec b}; y)$ is the gluon dipole scattering amplitude on the target nucleus, where we have also assumed that $Q_{s,2}^2(\vec b) \approx Q_{s,2}^2(\vec b +\vec r)$.

In order to write a $k_T$-factorized form for the two-gluon production cross section one must introduce new distribution functions that come with an extra transverse coordinate dependence and various Wilson line objects.
We have the unintegrated gluon distribution functions with extra coordinate dependence associated with the projectile
\begin{align}
\label{eq:dipole_wave} 
\left\langle \frac{d \phi_{A_1} ({\vec q}; y)}{d^2 b} \right\rangle_{A_1} =
  \frac{C_F}{\alpha_s ( 2 \pi)^3} \int d^2 r \; 
  e^{-i {\vec q} \cdot {\vec r}} \; \nabla_{{\vec r}}^2 \; n_G ({\vec b} + {\vec r}, {\vec b}; y)
\end{align}
 and target
\begin{align}
\label{eq:dipole_target} 
\left\langle \frac{d \phi_{A_2}({\vec q}; y) }{d^2 b} \right\rangle_{A_2} =
\frac{C_F}{\alpha_s ( 2 \pi)^3} \int d^2 r \; 
  e^{-i {\vec q} \cdot {\vec r}} \; \nabla_{{\vec r}}^2 \; N ({\vec b} + {\vec r}, {\vec b};y).
\end{align}
Here the coordinate dependence of the distribution functions only affects where   the saturation scale is measured, so for a translationally invariant nucleus the saturation scale becomes a constant and this coordinate dependence is absent.
Then, one finds that the distribution function above can be written in terms of unintegrated gluon distribution functions divided by the transverse area of the corresponding hadron or ion, $S_{\perp, 1}$ for the projectile and $S_{\perp, 2}$ for the target,
\begin{align}
\label{eq:dipole_wave} 
\left\langle \frac{d \phi_{A_i} ({\vec q}; y)}{d^2 b} \right\rangle_{A_i} =
  \frac{1}{S_{\perp, i}} \phi_{A_i} ({\vec q}; y).
\end{align}

In the $k_T$-factorized form, the interaction with the target can be described through two different  distribution functions. We have the double trace distribution
\begin{align} 
\label{eq:doubletrace_dist} 
\left\langle \frac{d \phi_{A_2}^{D} ({\vec  q}_1, {\vec  q}_2; y)}{d^2 b_1
  \; d^2 b_2} \right\rangle_{A_2} & = \left( \frac{C_F}{\alpha_s ( 2
  \pi)^3} \right)^2 \int d^2 r_1 \; d^2 r_2 \, e^{-i {\vec q}_1 \cdot
  {\vec r}_1 -i {\vec q}_2 \cdot {\vec r}_2}
\\ \notag
& \times \nabla_{{\vec r}_1}^2 \;
\nabla_{{\vec r}_2}^2 \; N_D ({\vec b}_1 + {\vec r}_1, {\vec b}_1, {\vec
  b}_2 + {\vec r}_2, {\vec b}_2; y),
\end{align} 
where
\begin{align} 
\label{eq:doubletrace_amp} 
N_D ({\vec x}, {\vec y}, {\vec z}, {\vec w}; y) = \; \frac{1}{(N_c^2-1)^2}
  \; \left\langle \mbox{Tr} \left[ \mathds{1} - U_{{\vec x}} U_{{\vec
  y}}^\dagger \right] \mbox{Tr} \left[ \mathds{1} - U_{{\vec z}}
  U_{{\vec w}}^\dagger \right] \right\rangle_{A_2} (y).
\end{align}
$U_{\vec{x}}$ is the Wilson line in the adjoint representation,
\begin{equation}
\label{eq:Wilson}
U_{\vec{x}}={\cal P}\exp \left[i g \int_{-\infty}^{+\infty} dx^+ {\cal A}^- (x^+,x^-=0,\vec{x}) \right],
\end{equation}
where $ {\cal A}$ is the gauge field of the hadron or nucleus in the adjoint representation, $x^-=0$ is the position of the infinitely contracted target nucleus (the shockwave), the light-cone gauge $ {\cal A}^+=0$ is used and quantities with arrows ($\vec{x},\vec{y},\dots$) are two-dimensional vectors that denote the transverse position.

Type 1 diagrams also contain the quadrupole distribution
\begin{align} 
\label{eq:quad_dist} 
\left\langle \frac{d \phi_{A_2}^{Q} ({\vec q}_1, {\vec q}_2; y)}{d^2 b_1
  \; d^2 b_2} \right\rangle_{A_2} & = \left( \frac{C_F}{\alpha_s ( 2
  \pi)^3} \right)^2 \int d^2 r_1 \; d^2 r_2 \, e^{-i {\vec q}_1 \cdot
  {\vec r}_1 -i {\vec q}_2 \cdot {\vec r}_2} \\ \notag
& \times \nabla_{{\vec r}_1}^2 \;
  \nabla_{{\vec r}_2}^2 \; N_{Q} ({\vec b}_1 + {\vec r}_1, {\vec b}_1, {\vec
  b}_2 + {\vec r}_2, {\vec b}_2; y),
\end{align}
with
\begin{align} 
\label{eq:quad_amp} 
N_{Q} ({\vec x}, {\vec y}, {\vec z}, {\vec w}; y) = \; \frac{1}{N_c^2-1}
  \; \left\langle \mbox{Tr} \left[ \left( \mathds{1} - U_{{\vec x}}
  U_{{\vec y}}^\dagger \right) \left( \mathds{1} - U_{{\vec z}}
  U_{{\vec w}}^\dagger \right) \right] \right\rangle_{A_2} (y).
\end{align}

Using these distribution functions we can write the cross section for two gluon production in dilute-dense collisions as the convolution of these various distribution functions and a kinematic kernel (see \cite{Kovchegov:2013ewa} for details),
\begin{align} 
\label{eq:factorized_final} 
& \frac{d \sigma}{d^2 k_1 dy_1 d^2 k_2 dy_2} = \left( \frac{2 \;
  \alpha_s}{C_F} \right)^2 \frac{1}{k_1^2 \; k_2^2} \int d^2 B \; d^2 b_1
  \; d^2 b_2 \int d^2 q_1 \; d^2 q_2  \\
& \times \left\langle \frac{d \phi_{A_1}
  ({\vec q}_1; y=0)}{d^2 ({\vec B}-{\vec b}_1)} \right\rangle_{\! \!
  A_1} \left\langle \frac{d \phi_{A_1} ({\vec q}_2; y=0)}{d^2 ({\vec
  B}-{\vec b}_2)} \right\rangle_{\! \! A_1}
  \left\{ \left\langle \frac{d \phi_{A_2}^{D} ({\vec q}_1 - {\vec k}_1,
  {\vec q}_2 - {\vec k}_2,;y)}{d^2 b_1 \; d^2 b_2} \right\rangle_{\!
  \! A_2} \! \! \right. \notag \\
& \left. + \, 
 e^{-i \, ( {\vec k}_1 - {\vec k}_2 )
  \cdot ( {\vec b}_1 - {\vec b}_2 )} \; \frac{\mathcal{K} (
  {\vec k}_1, {\vec k}_2, {\vec q}_1, {\vec q}_2)}{N_c^2-1}
  \left\langle \frac{d \phi_{A_2}^{Q} ({\vec q}_1 - {\vec k}_1, {\vec
  q}_2 - {\vec k}_2; y)}{d^2 b_1 \; d^2 b_2} \right\rangle_{\!
  \! A_2}
\right\} + ({\vec k}_2 \rightarrow - {\vec k}_2), \notag
\end{align}
with $\vec{B}$ the relative impact parameter between the target and the projectile (which is considered to be homogeneous and small compared with the target, consistent with the fact that gluons in the projectile should lie close enough to be correlated), and the kernel
\begin{align} 
\label{eq:kernel_crossed}
\mathcal{K} ({\vec k}_1, {\vec k}_2, {\vec q}_1,
  {\vec q}_2) & = \frac{1}{q_1^2 \; q_2^2 \; ({\vec k}_1-{\vec q}_1)^2
  ({\vec k}_2-{\vec q}_2)^2} \; \left\{ k_1^2 \; k_2^2 ({\vec
  q}_1 \cdot {\vec q}_2)^2 \right.  \notag \\
& - \; k_1^2 \; ({\vec
  q}_1 \cdot {\vec q}_2) \left[ ({\vec k}_2 \cdot {\vec q}_1) \; q_2^2
  \; + \; ({\vec k}_2 \cdot {\vec q}_2) \; q_1^2 \; - \; q_1^2 \;
  q_2^2 \right] \notag \\ & - \; k_2^2 \; ({\vec q}_1 \cdot {\vec
  q}_2) \left[ ({\vec k}_1 \cdot {\vec q}_1) \; q_2^2 \; + \; ({\vec
  k}_1 \cdot {\vec q}_2) \; q_1^2 \; - \; q_1^2 \; q_2^2 \right]
  \notag \\
& \left.  + \; q_1^2 \; q_2^2 \; \left[ ({\vec k}_1 \cdot
  {\vec q}_1) ({\vec k}_2 \cdot {\vec q}_2) \; + \; ({\vec k}_1 \cdot
  {\vec q}_2) ({\vec k}_2 \cdot {\vec q}_1) \right] \right\}.
\end{align}
In the next Section we will examine this expression and isolate the various contributions.


\section{Isolating the Various Contributions}
\label{section3}

The first contribution to the two gluon production cross section beyond the dilute-dilute limit (the glasma graph approximation) that one can isolate trivially is the classical contribution which stems from treating the produced gluons as two distinguishable particles. Effectively, this corresponds to ignoring all the interference diagrams  and considering only the diagrams of  type 3. Moreover, in this contribution the emissions of the two gluons are completely independent of each other. 

The correlations encoded in the classical contribution only depends on the geometry of the collision. Thus, it corresponds to uncorrelated production and it is the leading contribution in the large-$N_c$ limit (as discussed in \cite{Altinoluk:2015uaa}). Therefore, one can isolate this contribution by taking the large-$N_c$ limit of the two gluon production cross section, Eq.\eqref{eq:factorized_final}. In this limit, the double dipole operator given in Eq.\eqref{eq:doubletrace_amp} can be approximated as
\begin{eqnarray}
N_D ({\vec x}, {\vec y}, {\vec z}, {\vec w}; y) &\simeq& \; \frac{1}{N_c^2-1}
  \; \left\langle \mbox{Tr} \left[ \mathds{1} - U_{{\vec x}} U_{{\vec
  y}}^\dagger \right]\right\rangle_{A_2}(y) \;  \frac{1}{N_c^2-1} 
  \left\langle
  \mbox{Tr} \left[ \mathds{1} - U_{{\vec z}}
  U_{{\vec w}}^\dagger \right] \right\rangle_{A_2} (y)\nonumber\\
  &=& N_D ({\vec x}, {\vec y}; y) \; N_D ( {\vec z}, {\vec w}; y).
\end{eqnarray}
We would like to emphasize that this is not the only contribution containing the product of two target averaged dipoles but it is the only one that gives the uncorrelated piece and, moreover, the others are suppressed by factors $\frac{1}{(N_c^2-1)^2}$.
This corresponds to approximating the double trace distribution in the two gluon production cross section as 
\begin{equation}
\label{eq:Factorized_DD}
\left\langle \frac{d \phi_{A_2}^{D} ({\vec  q}_1-{\vec k}_1, {\vec  q}_2-{\vec k_2}; y)}{d^2 b_1
  \; d^2 b_2} \right\rangle_{A_2} \simeq 
  \left\langle \frac{d\phi_{A_2}({\vec q}_1-{\vec k_1}; y)}{d^2b_1}\right\rangle_{A_2} \; 
   \left\langle \frac{d\phi_{A_2}({\vec q}_2-{\vec k_2}; y)}{d^2b_2}\right\rangle_{A_2} \; .
\end{equation} 
Finally, the classical contribution to the two-gluon production cross section reads
\begin{align} 
\label{eq:factorized_geo} 
& \frac{d \sigma_{classical}}{d^2 k_1 dy_1 d^2 k_2 dy_2} = \left( \frac{2 \;
  \alpha_s}{C_F} \right)^2 \frac{1}{k_1^2 \; k_2^2} \int d^2 B \; d^2 b_1
  \; d^2 b_2 \int d^2 q_1 \; d^2 q_2  \\
& \times \left\langle \frac{d \phi_{A_1}
  ({\vec q}_1; y=0)}{d^2 ({\vec B}-{\vec b}_1)} \right\rangle_{\! \!
  A_1} \left\langle \frac{d \phi_{A_1} ({\vec q}_2; y=0)}{d^2 ({\vec
  B}-{\vec b}_2)} \right\rangle_{\! \! A_1}
  \left\langle \frac{d \phi_{A_2}
  ({\vec q}_1 -{\vec k}_1; y)}{d^2 b_1} \right\rangle_{\! \!
  A_2}\left\langle \frac{d \phi_{A_2}
  ({\vec q}_2-{\vec k}_2; y)}{d^2 b_2} \right\rangle_{\! \!
  A_2}.\nonumber
\end{align}
As it was mentioned previously, the only correlations contained in the classical contribution are the ones that are related with the geometry of the collision. However, these correlations (that are not of particular interest for this study) can be neglected assuming a translationally invariant target. In this case, integrations over various impact parameters can be performed trivially and the classical contribution to the two gluon production cross section reads 
\begin{align} 
\label{eq:classical} 
& \frac{d \sigma_{classical}}{d^2 k_1 dy_1 d^2 k_2 dy_2} = \frac{1}{S_{\perp, 2}} \left( \frac{2 \;
  \alpha_s}{C_F} \right)^2 \frac{1}{k_1^2 \; k_2^2} \int d^2 q_1 \; d^2 q_2 \notag \\
& \times \phi_{A_1} ({\vec q}_1; y=0) \; \phi_{A_1} ({\vec q}_2; y=0) \;
  \phi_{A_2} ({\vec q}_1 -{\vec k}_1; y) \; \phi_{A_2} ({\vec q}_2 -{\vec k}_2; y)
\end{align}
which is just the square of the single-gluon production cross section divided by the transverse area, i.e.,
\begin{equation}
\frac{d \sigma_{classical}}{d^2 k_1 dy_1 d^2 k_2 dy_2} =\frac{1}{S_{\perp, 2}} \frac{d \sigma_{g}}{d^2 k_1 dy_1} \frac{d \sigma_{g}}{d^2 k_2 dy_2}.
\end{equation}

Our next order of business is to isolate the HBT and Bose enhancement contributions. As it was discussed in detail in \cite{Altinoluk:2015uaa} and \cite{Altinoluk:2015eka}, these contributions are suppressed by a factor of $\frac{1}{N_c^2-1}$ when compared to the classical contribution and they can originate either from the quadrupole distribution or the double dipole distribution terms in the two gluon production cross section, Eq.\eqref{eq:factorized_final}.

First, we consider the quadrupole distribution and factorize the target averaging of the four Wilson lines into averaging over the pairs as it was shown in \cite{Kovchegov:2013ewa}\footnote {In principle, a pair of gluons can be found in all of the seven irreducible representations of $SU(N_c)$ that result from the product of two adjoints. However, we have approximated the quadrupole operator by only considering the singlet projector which gives the factorized double dipole operator as it was argued in \cite{Kovchegov:2013ewa}.}. Then, the quadrupole term reads 
\begin{align}
& \left\langle 
\mbox{Tr} 
\left[ U_{{\vec r}_1+{\vec b}_1}  U^{\dagger}_{{\vec b}_1}  U_{{\vec r}_2+{\vec b}_2} U^{\dagger}_{{\vec b}_2} \right]
\right\rangle_{A_2} (y)
\\
& \quad \quad \quad =
\frac{1}{N_c^2-1}
\left\langle 
\mbox{Tr} \left[ U_{{\vec r}_1+{\vec b}_1} U^{\dagger}_{{\vec b}_1} \right]
\right\rangle_{A_2} (y)
\left\langle 
\mbox{Tr} \left[ U_{{\vec r}_2+{\vec b}_2} U^{\dagger}_{{\vec b}_2} \right]
\right\rangle_{A_2} (y) \nonumber\\
& \quad \quad \quad +
\frac{1}{N_c^2-1}
\left\langle 
\mbox{Tr} \left[ U_{{\vec r}_1+{\vec b}_1} U^{\dagger}_{{\vec b}_2} \right]
\right\rangle_{A_2} (y)
\left\langle 
\mbox{Tr} \left[ U_{{\vec r}_2+{\vec b}_2} U^{\dagger}_{{\vec b}_1} \right]
\right\rangle_{A_2} (y) \nonumber \\
& \quad \quad \quad +
\frac{1}{(N_c^2-1)^2}
\left\langle 
\mbox{Tr} \left[ U_{{\vec r}_1+{\vec b}_1} U^{\dagger}_{{\vec r}_2+{\vec b}_2} \right]
\right\rangle_{A_2} (y)
\left\langle 
\mbox{Tr} \left[ U_{{\vec b}_1} U^{\dagger}_{{\vec b}_2} \right]
\right\rangle_{A_2} (y) \nonumber \\
& \quad \quad \quad + \cdots \ .\nonumber
\end{align}
Using this factorization, the gradients act on related dipoles and the quadrupole operator can be written in terms of the dipole operators as
\begin{eqnarray}
\label{eq:Factorized_Quad}
\nabla^2_{{\vec r}_1} \nabla^2_{{\vec r}_2}N_Q({\vec r}_1+{\vec b}_1, {\vec b}_1, {\vec r}_2+{\vec b}_2, {\vec b}_2; y)
&=&
\nabla^2_{{\vec r}_1} N ({\vec r}_1+{\vec b}_1, {\vec b}_1; y) 
\nabla^2_{{\vec r}_2} N ({\vec r}_2+{\vec b}_2, {\vec b}_2; y) \\
&+&
\nabla^2_{{\vec r}_1} N({\vec r}_1+{\vec b}_1, {\vec b}_2; y)
\nabla^2_{{\vec r}_2} N({\vec r}_2+{\vec b}_2, {\vec b}_1; y)\nonumber\\
&+&
\frac{1}{N_c^2-1}
\nabla^2_{{\vec r}_1} \nabla^2_{{\vec r}_2} N ({\vec r}_1+{\vec b}_1, {\vec r}_2+{\vec b}_2; y) N({\vec b}_1, {\vec b}_2; y) \nonumber \\
&+& \cdots \ .\nonumber
\end{eqnarray}  
One can use this factorized form of the quadrupole operator, Eq. \eqref{eq:Factorized_Quad}, to get the explicit expressions for the quadrupole distributions of the target in terms of the dipole distributions which then will be used to identify various contributions to the two gluon production cross section. 

Once the first term of Eq. \eqref{eq:Factorized_Quad} is substituted into the quadrupole distribution, one gets
\begin{eqnarray}
\left\langle \frac{\phi_{term \, 1}({\vec q}_1-{\vec k}_1,{\vec q}_2-{\vec k}_2; y)}{d^2b_1d^2b_2}\right\rangle_{A_2}
= \left( \frac{C_F}{\alpha_s(2\pi)^3}\right)^2
\int d^2r_1 d^2 r_2 e^{-i{\vec r}_1\cdot({\vec q}_1-{\vec k}_1)-i{\vec r}_2\cdot({\vec q}_2-{\vec k}_2)}\nonumber\\
\times 
\nabla^2_{{\vec r}_1} N({\vec r}_1+{\vec b}_1,{\vec b}_1 ; y)
\nabla^2_{{\vec r}_2} N({\vec r}_1+{\vec b}_2, {\vec b}_2; y)
\nonumber\\
=
\left\langle \frac {d\phi_{A_2}({\vec q}_1-{\vec k}_1; y)}{d^2b_1} \right\rangle_{A_2}
\left\langle \frac {d\phi_{A_2}({\vec q}_2-{\vec k}_2; y)}{d^2b_2} \right\rangle_{A_2} \, .
\end{eqnarray}
Now, this distribution can be plugged into the two gluon production cross section, Eq. \eqref{eq:factorized_final}, to get the contribution of the first term of Eq. \eqref{eq:Factorized_Quad}  which simply reads
\begin{align} 
\label{eq:factorized_HBT} 
& \frac{d \sigma_{term \, 1}}{d^2 k_1 dy_1 d^2 k_2 dy_2} = \left( \frac{2 \;
  \alpha_s}{C_F} \right)^2 \frac{1}{k_1^2 \; k_2^2} \int d^2 B \; d^2 b_1
  \; d^2 b_2 \int d^2 q_1 \; d^2 q_2 \notag \\
& \times
  \left\langle \frac{d \phi_{A_1} ({\vec q}_1;y=0)}{d^2 ({\vec B}-{\vec b}_1)} \right\rangle_{\! \! A_1}
  \left\langle \frac{d \phi_{A_1} ({\vec q}_2;y=0)}{d^2 ({\vec B}-{\vec b}_2)} \right\rangle_{\! \! A_1}
  \left\langle \frac{d \phi_{A_2} ({\vec q_1}-{\vec k_1};y)}{d^2 b_1} \right\rangle_{A_2}
  \left\langle \frac{d \phi_{A_2} ({\vec q_2}-{\vec k_2};y)}{d^2 b_2} \right\rangle_{A_2}
\notag \\
& \times \left[ e^{-i \, ( {\vec k}_1 - {\vec k}_2 )
  \cdot ( {\vec b}_1 - {\vec b}_2 )} \; \frac{\mathcal{K} (
  {\vec k}_1, {\vec k}_2, {\vec q}_1, {\vec q}_2)}{N_c^2-1}
  \right] + ({\vec k}_2 \rightarrow - {\vec k}_2)   \, .
\end{align}
This term was identified as the HBT contribution in \cite{Kovchegov:2013ewa} and in \cite{Altinoluk:2015eka}. It is easier to understand this identification in the case of a translationally invariant target:
\begin{align} 
\label{eq:HBTtrans} 
& \frac{d \sigma_{HBT}}{d^2 k_1 dy_1 d^2 k_2 dy_2} = \frac{1}{S_{\perp, 1} S_{\perp, 2}} \left( \frac{2 \;
  \alpha_s}{C_F} \right)^2 \frac{1}{k_1^2 \; k_2^2} \int d^2 q_1 \; d^2 q_2 \notag \\
& \times
  \phi_{A_1} ({\vec q}_1;y=0) \; \phi_{A_1} ({\vec q}_2;y=0) \;
  \phi_{A_2} ({\vec q}_1 - \vec k_1;y) \; \phi_{A_2} ({\vec q}_2 - \vec k_2;y)
\notag \\
& \times  \frac{2 \pi }{N_c^2-1}
  \left( \delta(\vec k_1 - \vec k_2) + \delta(\vec k_1 + \vec k_2) \right)
  \mathcal{K} ({\vec k}_1, {\vec k}_2, {\vec q}_1, {\vec q}_2)
\end{align}
which clearly gives a peak at ${\vec k}_1={\vec k}_2$ and at ${\vec k}_1=-{\vec k}_2$ (as expected from the HBT contribution), with ${\vec k}_1$ and ${\vec k}_2$ being the transverse momenta of the produced gluons. In order to observe the relative enhancement of the HBT contribution with respect to the Bose-enhanced one, we should consider the origin of the $\delta$-functions in \eqref{eq:HBTtrans}. They come from an integral in $\vec b_1-\vec{b_2}$ and thus, in the non-translational invariant case, provide a factor $S_{\perp, 1}$ that enhance this contribution by the number of sources $S_{\perp, 1}Q^2_{s,1}$ with respect to the Bose-enhanced term that we discuss next.

In a similar manner, we can consider the second term of Eq. \eqref{eq:Factorized_Quad}. When substituted in the quadrupole distribution, it reads 
\begin{eqnarray}
\left\langle \frac{\phi_{term \, 2}({\vec q}_1-{\vec k}_1,{\vec q}_2-{\vec k}_2; y)}{d^2b_1d^2b_2}\right\rangle_{A_2}
= \left( \frac{C_F}{\alpha_s(2\pi)^3}\right)^2
\int d^2r_1 d^2 r_2 e^{-i{\vec r}_1\cdot({\vec q}_1-{\vec k}_1)-i{\vec r}_2\cdot({\vec q}_2-{\vec k}_2)}\nonumber\\
\times 
\nabla^2_{{\vec r}_1} N({\vec r}_1+\Delta{\vec b}+{\vec b}_2, {\vec b}_2 ; y)
\nabla^2_{{\vec r}_2} N({\vec r}_1-\Delta{\vec b}+{\vec b}_1,{\vec b}_1 ; y),
\end{eqnarray}
where we have introduced $\Delta{\vec b}={\vec b}_1-{\vec b}_2$. It is convenient to define the shifted variables
\begin{eqnarray}
{\vec r}\,'_1&=&{\vec r}_1+\Delta{\vec b}, \nonumber\\
{\vec r}\,'_2&=&{\vec r}_2-\Delta{\vec b},
\end{eqnarray}  
in order to write the quadrupole distribution of the second term of  Eq. \eqref{eq:Factorized_Quad} in factorized form as 
\begin{eqnarray}
&&\left\langle \frac{\phi_{term \, 2}({\vec q}_1-{\vec k}_1,{\vec q}_2-{\vec k}_2; y)}{d^2b_1d^2b_2}\right\rangle_{A_2}
= \left( \frac{C_F}{\alpha_s(2\pi)^3}\right)^2
e^{-i(\Delta{\vec b})\cdot({\vec q}_2-{\vec k}_2-{\vec q}_1+{\vec k}_1)}\nonumber\\
&&\times
\int d^2r'_1 d^2 r'_2 e^{-i{\vec r}\,'_1\cdot({\vec q}_1-{\vec k}_1)-i{\vec r}\,'_2\cdot({\vec q}_2-{\vec k}_2)}
\nabla^2_{{\vec r}\,'_1} N({\vec r}\,'_1+{\vec b}_2, {\vec b}_2 ; y)
\nabla^2_{{\vec r}\,'_2} N({\vec r}\,'_2+{\vec b}_1,{\vec b}_1 ; y)\nonumber\\
&&=\,
e^{-i(\Delta{\vec b})\cdot({\vec q}_2-{\vec k}_2-{\vec q}_1+{\vec k}_1)}
 \left\langle \frac{d \phi_{A_2} ({\vec q_1}-{\vec k_1};y)}{d^2 b_2} \right\rangle_{A_2}
  \left\langle \frac{d \phi_{A_2} ({\vec q_2}-{\vec k_2};y)}{d^2 b_1} \right\rangle_{A_2}\ \! \! \! .
\end{eqnarray}
We plug this factorized form of the second term of the quadrupole distribution into the two gluon production cross section and the result reads
\begin{align} 
\label{eq:factorized_Bose} 
& \frac{d \sigma_{term \, 2}}{d^2 k_1 dy_1 d^2 k_2 dy_2} = \left( \frac{2 \;
  \alpha_s}{C_F} \right)^2 \frac{1}{k_1^2 \; k_2^2} \int d^2 B \; d^2 b_1
  \; d^2 b_2 \int d^2 q_1 \; d^2 q_2 \notag \\
& \times
  \left\langle \frac{d \phi_{A_1} ({\vec q}_1;y=0)}{d^2 ({\vec B}-{\vec b}_1)} \right\rangle_{\! \! A_1}
  \left\langle \frac{d \phi_{A_1} ({\vec q}_2;y=0)}{d^2 ({\vec B}-{\vec b}_2)} \right\rangle_{\! \! A_1}
  \left\langle \frac{d \phi_{A_2} ({\vec q_1}-{\vec k_1};y)}{d^2 b_2} \right\rangle_{A_2}
  \left\langle \frac{d \phi_{A_2} ({\vec q_2}-{\vec k_2};y)}{d^2 b_1} \right\rangle_{A_2}
\notag \\
& \times 
\frac{1}{N_c^2-1} \, 
e^{-i{\Delta{\vec b}}\cdot(2{\vec k}_1-2{\vec k}_2-{\vec q}_1+{\vec q}_2)}
\;   \mathcal{K} ({\vec k}_1, {\vec k}_2, {\vec q}_1, {\vec q}_2) 
\;   + \; ({\vec k}_2 \rightarrow - {\vec k}_2) .
\end{align}
For convenience, let us define the average transverse momentum ${\vec q}$ and shifted transverse momenta difference $\Delta{\vec q}$ as
\begin{eqnarray}
{\vec q}&=&\frac{1}{2}({\vec q}_1+{\vec q}_2),\\
\Delta{\vec q}&=&{\vec q}_1-{\vec q}_2-2\Delta{\vec k},\nonumber
\end{eqnarray}
where we have defined the transverse momenta difference of the produced gluons as $\Delta{\vec k}={\vec k}_1-{\vec k}_2$. After these change of variables the contribution of the second term of the quadrupole distribution to the two gluon production cross section can be written as 
\begin{align} 
& \frac{d \sigma_{term \, 2}}{d^2 k_1 dy_1 d^2 k_2 dy_2} = \left( \frac{2 \;
  \alpha_s}{C_F} \right)^2 \frac{1}{k_1^2 \; k_2^2} \int d^2 B \; d^2 b_1
  \; d^2 b_2 \int d^2 q \; d^2 \Delta q    
  \; \frac{1}{N_c^2-1} \, 
e^{i{\Delta{\vec b}}\cdot\Delta{\vec q}}    \notag \\
& \times
  \left\langle \frac{d \phi_{A_1} ({\vec q}+\Delta{\vec q}/2-\Delta{\vec k};y=0)}{d^2 ({\vec B}-{\vec b}_1)} \right\rangle_{\! \! A_1}
  \left\langle \frac{d \phi_{A_1} ({\vec q}-\Delta{\vec q}/2-\Delta{\vec k};y=0)}{d^2 ({\vec B}-{\vec b}_2)} \right\rangle_{\! \! A_1}
  \nonumber\\
  &
  \times
  \left\langle \frac{d \phi_{A_2} ({\vec q}-\Delta{\vec q}/2-{\vec k}_1;y)}{d^2 b_1} \right\rangle_{A_2}
  \left\langle \frac{d \phi_{A_2} ({\vec q}+\Delta{\vec q}/2-{\vec k}_2;y)}{d^2 b_2} \right\rangle_{A_2}
\notag \\
& \times 
\;   \mathcal{K} \left({\vec k}_1, {\vec k}_2, {\vec q}+\frac{\Delta{\vec q}}{2}+\Delta{\vec k}, {\vec q}-\frac{\Delta{\vec q}}{2}-\Delta{\vec k}\right) 
\;   + \; ({\vec k}_2 \rightarrow - {\vec k}_2) \; .
\end{align}
In the case of a translationally invariant target, this contribution reads
\begin{align} 
\label{eq:Bose_final} 
& \frac{d \sigma_{Bose}}{d^2 k_1 dy_1 d^2 k_2 dy_2} = \left( \frac{2 \;
  \alpha_s}{C_F} \right)^2 \frac{1}{k_1^2 \; k_2^2} \; \frac{1}{S_{\perp, 1} S_{\perp, 2}} \; \frac{1}{N_c^2-1}
   \int d^2q \; 
   \mathcal{K}\left( {\vec k}_1,{\vec k}_2, {\vec q}+\Delta{\vec k},{\vec q}-\Delta{\vec k}\right) \notag  \\
   &
   \times
   \phi_{A_1}({\vec q}+\Delta{\vec k}; y=0) \,  \phi_{A_1}({\vec q}-\Delta{\vec k}; y=0) \, 
   \phi_{A_2}({\vec q}-{\vec k}_1; y) \,  \phi_{A_2}({\vec q}-{\vec k}_2; y) \,  \notag  \\
   & + \; ({\vec k}_2 \rightarrow - {\vec k}_2) \; .
\end{align}
We have identified this term as Bose enhancement of the projectile even though the projectile gluon distributions seem to have different momenta.  However, as discussed in detail in \cite{Altinoluk:2015uaa}, the Bose enhancement contribution to the correlated production is peaked when the ${\vec k}_1={\vec k}_2$ for the nearside and when ${\vec k}_1=-{\vec k}_2$ for the away side ridge. Thus, for the near side ridge, $\Delta{\vec k}\to 0$ and clearly we get the peak in the first term of Eq. \eqref{eq:Bose_final}. Note that, the same argument holds for the away side ridge and the second term of Eq. \eqref{eq:Bose_final}. In conclusion, the Bose enhancement contribution for the near side ridge reads
\begin{align}
\label{eq:Bose_near} 
& \frac{d \sigma_{Bose}}{d^2 k_1 dy_1 d^2 k_2 dy_2}\bigg|_{{\vec k}_1={\vec k}_2={\vec k}} = \left( \frac{2 \;
  \alpha_s}{C_F} \right)^2 \frac{1}{k^4} \; \frac{1}{S_{\perp, 1} S_{\perp, 2}} \; \frac{1}{N_c^2-1}
   \int d^2q \; 
   \mathcal{K}\left( {\vec k},{\vec k}, {\vec q},{\vec q}\right)  \\
   & \hspace{4cm}
   \times
   \phi_{A_1}({\vec q}; y=0) \,  \phi_{A_1}({\vec q}; y=0) \, 
   \phi_{A_2}({\vec q}-{\vec k}; y) \,  \phi_{A_2}({\vec q}-{\vec k}; y),  \; 
   \notag 
\end{align}
with the kernel $ \mathcal{K}\left( {\vec k},{\vec k}, {\vec q},{\vec q}\right)$ being 
\begin{equation}
\mathcal{K}\left( {\vec k},{\vec k}, {\vec q},{\vec q}\right)= \bigg\{ 1+\frac{1}{(|{\vec k}-{\vec q}|^2)^2}
\Big[ 4q^2({\vec k}\cdot{\vec q}) - |q^2|^2 + 2({\vec k}\cdot{\vec q})^2 \Big]\bigg\} \; .
\end{equation}

We have identified the HBT, Eq.  \eqref{eq:HBTtrans}, and the Bose enhancement, Eq. \eqref{eq:Bose_final}, contributions to the correlated production that stem respectively from the first and second terms of the quadrupole distributions given in Eq. \eqref{eq:Factorized_Quad}. Besides, we have also identified the classical contribution, Eq. \eqref{eq:classical} that originates from the double dipole term, Eq. \eqref{eq:Factorized_DD}. It contributes to  uncorrelated production and it is the leading term in the large $N_c$ limit. These are the main results of the paper. However, we would like to comment about the remaining terms of the quadrupole and the double dipole distributions. It is clear from the last line of Eq. \eqref{eq:Factorized_Quad} that the third contribution of the quadrupole distribution is suppressed by an extra power of $\frac{1}{N_c^2-1}$ with respect to the first two contributions that are responsible for HBT and Bose enhancement contributions.  On the other hand, the remaining two contributions of the double dipole distribution are suppressed an extra power of $\frac{1}{N_c^2-1}$ as well \footnote{This can be easily understood from the fact that two color projectors, one on the right and one on the left, are required to express the double dipole term with a combination of coordinates that would result in Bose enhancement, see \cite{Kovchegov:2013ewa}.}.
Therefore, the complete results for the uncorrelated, HBT and Bose-enhanced pieces of the two-gluon production cross section in the dilute-dense limit  correspond to Eqs. \eqref{eq:classical}, \eqref{eq:HBTtrans} and \eqref{eq:Bose_final} for the translational invariant case.


\section{A toy model}
\label{toymodel}

In this section, we perform the numerical analysis of the main results of our study, namely the HBT and the Bose enhancement contributions to the correlated two-gluon production, by adopting a toy model for both the projectile and the target distributions.  

The unintegrated gluon distributions of the projectile are defined in Eq. \eqref{eq:unint_wave} with $n_G ({\vec b} + {\vec r}, {\vec b};y)$ being the distribution associated with the dilute projectile, whose expression is given in Eq. \eqref{project_dis}. In our toy model, we assume translational invariance of the dilute projectile. Effectively, this is equivalent to approximate the saturation scale of the projectile that depends on the impact parameter $b$ by  a constant which serves as an infrared cut off, i.e. $Q^2_{s,1}({\vec b})\approx Q^2_1$. Within the limits of this  approximation, the unintegrated gluon distribution of the projectile can be written as 

\begin{equation}
\label{eq:projectile_distribution}
\phi_{A_1}({\vec q})\approx \frac{C_F \, S_{\perp, 1} \, Q_1^2 }{\alpha_s \, (2\pi)^3}\, \frac{1}{4} \int d^2 r \, e^{-i{\vec q}\cdot {\vec r}} \, \nabla^2_{\vec r}\left[ r^2 \mbox{ln}\left(\frac{1}{r \, \Lambda }\right)\right] = 
\frac{C_F \, S_{\perp, 1} \, Q_1^2 }{\alpha_s \, (2\pi)^3}\,  \frac{2\pi}{q^2}\ .
\end{equation} 

On the other hand, we adopt the Golec-Biernat--W\"usthoff (GBW) model \cite{GolecBiernat:1998js} for the dipole distribution of the target: 
\begin{equation}
N_G({\vec r}+{\vec b},{\vec b};y)=1-e^{-\frac{Q_2^2}{4}r^2},
\end{equation} 
with $Q_2$ being the saturation scale of the target. Then, the target distribution reads
\begin{eqnarray}
\label{eq:target_distribution}
\phi_{A_2}({\vec q})= \frac{C_F}{\alpha_s (2\pi)^3}\int d^2r d^2b \,e^{-i{\vec q}\cdot{\vec r}} \, \nabla^2_{{\vec r}}\left( 1-e^{-\frac{Q_2^2}{4}r^2}\right)
= \frac{C_F}{\alpha_s (2\pi)^3} \, S_{\perp, 2} \, \frac{q^2}{Q_2^2}\, 4\pi\, e^{-\frac{q^2}{Q_2^2}} .
\end{eqnarray}
Using the projectile distribution, Eq.\eqref{eq:projectile_distribution}, and the target distribution, Eq. \eqref{eq:target_distribution}, the HBT contribution to the two gluon production cross section, Eq. \eqref{eq:HBTtrans}, can be written as 

\begin{eqnarray}
\label{eq:HBT_modeled}
\frac{d\sigma_{HBT}}{d^2k_1dy_1d^2k_2dy_2}&=&
\left(\frac{C_F}{\alpha_s}\right)^2 16 \, \frac{1}{(2\pi)^8}\, S_{\perp, 1} S_{\perp, 2} \, \left(\frac{Q_1^2}{Q_2^2}\right)^2\, \frac{1}{k_1^2k_2^2} \, \int d^2q_1 d^2q_2
\frac{(\vec q_1-\vec k_1)^2(\vec q_2-\vec k_2)^2}{q_1^2q_2^2} \nonumber\\
&&
\hspace{-2cm}
\times \frac{2\pi}{N_c^2-1}
e^{-\frac{1}{Q_2^2}\left[ (\vec q_1-\vec k_1)^2+(\vec q_2-\vec k_2)^2\right]} \, \mathcal{K}(\vec k_1,\vec k_2,\vec q_1,\vec q_2) \, 
\left[ \delta^{(2)}(\vec k_1-\vec k_2) + \delta^{(2)}(\vec k_1+\vec k_2)\right] \, .
\end{eqnarray}

Similarly, the Bose enhancement contribution to the two gluon production cross section reads
\begin{eqnarray}
\label{eq:Bose_Modeled}
\frac{d\sigma_{Bose}}{d^2k_1dy_1d^2k_2dy_2}&=&\left(\frac{C_F}{\alpha_s}\right)^2 16 \, \frac{1}{(2\pi)^8}
\, S_{\perp, 1}S_{\perp, 2} \, \left(\frac{Q_1^2}{Q_2^2}\right) \, \frac{1}{k_1^2k_2^2}\notag \\
&&
\hspace{-3.5cm}
\times \int d^2q
\frac{\left( \vec q-\bar{k}-\frac{\Delta \vec k}{2}\right)^2 \left( \vec q-\bar{k}+\frac{\Delta \vec k}{2}\right)^2}{(\vec q+\Delta \vec k)^2(\vec q-\Delta \vec k)^2}\frac{2\pi}{N_c^2-1} \,\exp{\left\{-\frac{2}{Q_2^2}(\vec q-\bar{k})^2-\frac{1}{Q_2^2}(\Delta \vec k)^2\right\}}
\notag\\
&&
\hspace{-3.5cm}
\times \ 
\mathcal{K}(\vec k_1,\vec k_2,\vec q+\Delta \vec k, \vec q-\Delta \vec k) + ({\vec k}_2 \rightarrow - {\vec k}_2),
\end{eqnarray}
where $\bar{k}=(\vec k_1+\vec k_2)/2$.

\begin{figure}[htbp]
\centering
\includegraphics[width=\textwidth]{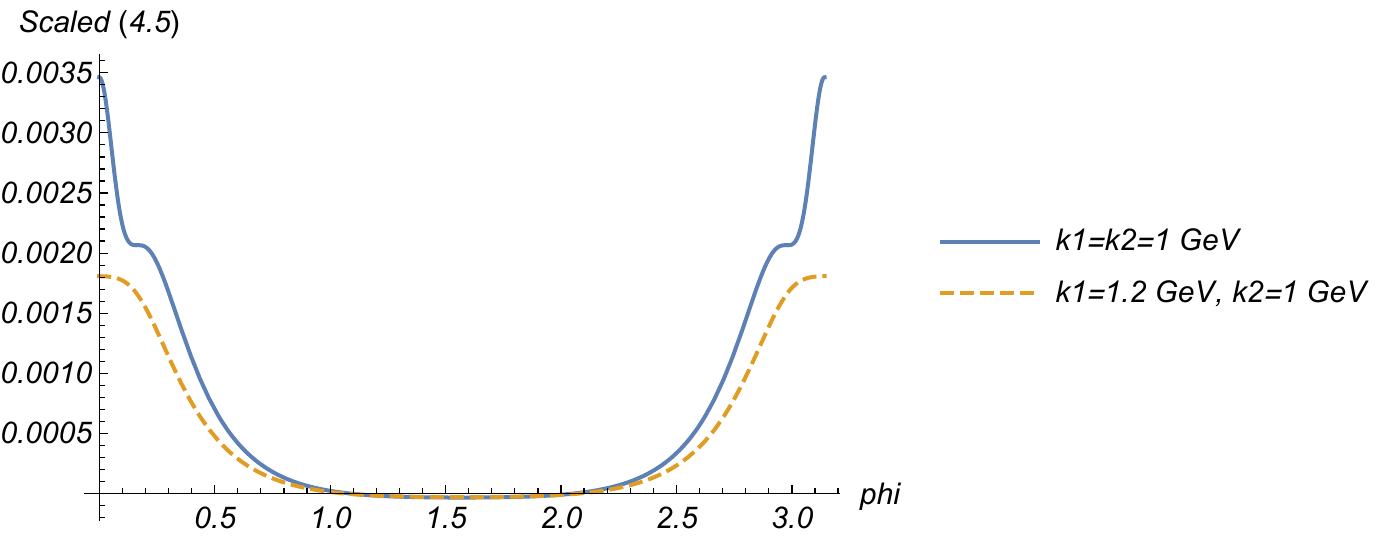}
\caption{\label{fig:3} Eq. \eqref{eq:Bose_Modeled} divided by $\left(\frac{C_F}{\alpha_s}\right)^2 16 \, \frac{1}{(2\pi)^8}
\,  S_{\perp, 1}S_{\perp, 2} \, \frac{1}{k_1^2k_2^2}$ for different choices of $k_1,k_2$ and $Q_1=0.2$ GeV and $Q_2=1$ GeV, versus the angle $\phi$ between them.}
\end{figure}

\begin{figure}[htbp]
\centering
\includegraphics[width=\textwidth]{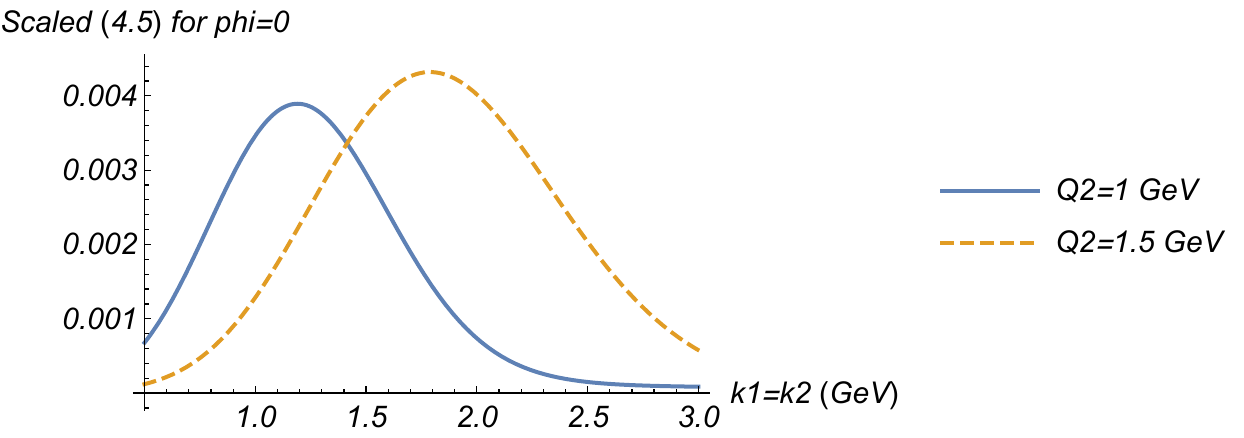}
\caption{\label{fig:4} Eq. \eqref{eq:Bose_Modeled} divided by $\left(\frac{C_F}{\alpha_s}\right)^2 16 \, \frac{1}{(2\pi)^8}
\,  S_{\perp, 1}S_{\perp, 2} \, \frac{1}{k_1^2k_2^2}$ for  $\phi=0$,  $Q_1=0.2$ GeV and different choices of $Q_2$, versus $k_1=k_2$.}
\end{figure}

In Figs. \ref{fig:3} and \ref{fig:4} we plot the result of Eq. \eqref{eq:Bose_Modeled} divided by $\left(\frac{C_F}{\alpha_s}\right)^2 16 \, \frac{1}{(2\pi)^8}
\, S_{\perp, 1}S_{\perp, 2}\, \frac{1}{k_1^2k_2^2}$ for different choices of $k_1,k_2$ and the angle $\phi$ between them, and $Q_1=0.2$ GeV and $Q_2=1$ GeV. The denominators in \eqref{eq:Bose_Modeled} resulting in divergent contributions have been regulated by adding them $Q_1^2$ or $Q_2^2$ if they stem from the projectile or target dipole distributions respectively.

While the model that we have used cannot be considered realistic, it illustrates several of the features of the result. First, in Fig. \ref{fig:3} the ridge structure can be seen, symmetric for the near and away side peaks in this calculation. The dip observed at $\phi\simeq 0.2, \pi-0.2$ comes from the double Gaussian structure in Eq. \eqref{eq:Bose_Modeled} in this model. Second, a fast degradation of the ridge for  $k_1 \ne k_2$ can also be observed, a clear signal of the effect of Bose statistics. Finally, in Fig. \ref{fig:4} the height of the peak for $k_1=k_2$ has a maximum for $k_1$ slightly above $Q_2$ and is seen to decrease very fast for larger $k_1$.


\section{Conclusions}
\label{conclusions}

In conclusion, in this paper we have explicitly considered the two-gluon production cross section in the CGC for the collision of a dilute projectile on a dense target, going beyond the glasma graph approximation that has been used in the dilute-dilute limit.  In this dilute-dense limit, applicable for proton-nucleus or heavy-light ion collisions, we have identified the HBT and the Bose enhancement contributions to the correlated production that are given in Eqs. \eqref{eq:HBTtrans} and \eqref{eq:Bose_final} respectively. The latter comes suppressed by the number of particles sources with respect to the HBT one, and shows the characteristic suppression by the number of degress of freedom with respect to the uncorrelated contribution.

We have shown that both contributions survive the inclusion of higher order density corrections and that they stem from the quadrupole distribution of the target. We have established the correspondence between the glasma graph approximation and the $k_T$-factorized approach, showing that these contributions come from type 1 and type 2 diagrams in the glasma graph approach that correspond to the interference diagrams in the $k_T$-factorized formulation. 
On the other hand,  we have also identified the classical contribution and have shown that it contributes to the uncorrelated production in the case of a translationally invariant target. We have identified that the origin of this contribution is the type 3 diagrams.
Finally, we have developed a toy model that allows a simple numerical implementation and whose results illustrate some of the features of the approach. 
The findings in this study are coherent with the results of the previous works \cite{Kovchegov:2012nd,Kovchegov:2013ewa,Altinoluk:2015uaa,Altinoluk:2015eka} on two-gluon production, and also with similar works \cite{Altinoluk:2016vax,Martinez:2018ygo,Kovner:2017vro, Kovner:2017ssr, Kovner:2018vec} on double and multi-quark production.

\acknowledgments
TA expresses his gratitude to the Departamento de F\'{\i}sica de Part\'{\i}culas at Universidade de Santiago de Compostela, for support and hospitality when part of this work was done. The work of TA is supported by Grant No. 2017/26/M/ST2/01074 of the National Science Centre, Poland. The work of NA and DEW were supported by the European Research Council grant HotLHC ERC-2011-StG-279579, Ministerio de Ciencia e Innovaci\'on of Spain under project FPA2014-58293-C2-1-P and Unidad de Excelencia Mar\'{\i}a de Maetzu under project MDM-2016-0692,  Xunta de Galicia (Conseller\'{\i}a de Educaci\'on) within the Strategic Unit AGRUP2015/11, and FEDER. 
This work has been performed in the framework of COST Action CA15213 "Theory of hot matter and relativistic heavy-ion collisions" (THOR).



\begin{thebibliography}{99}

\bibitem{Alver:2009id}
  B.~Alver {\it et al.} [PHOBOS Collaboration],
  Phys.\ Rev.\ Lett.\  {\bf 104} (2010) 062301
  [arXiv:0903.2811 [nucl-ex]].
  
\bibitem{Abelev:2009af}
  B.~I.~Abelev {\it et al.} [STAR Collaboration],
  Phys.\ Rev.\ C {\bf 80} (2009) 064912
  [arXiv:0909.0191 [nucl-ex]].

\bibitem{Khachatryan:2010gv}
  V.~Khachatryan {\it et al.} [CMS Collaboration],
  JHEP {\bf 1009} (2010) 091
  [arXiv:1009.4122 [hep-ex]].

\bibitem{Khachatryan:2015lva}
  Phys.\ Rev.\ Lett.\  {\bf 116}   (2016) 172302
  [arXiv:1510.03068 [nucl-ex]].

\bibitem{Aad:2015gqa}
  G.~Aad {\it et al.} [ATLAS Collaboration],
  Phys.\ Rev.\ Lett.\  {\bf 116}  (2016) 172301
  [arXiv:1509.04776 [hep-ex]].

\bibitem{CMS:2012qk}
  S.~Chatrchyan {\it et al.} [CMS Collaboration],
  Phys.\ Lett.\ B {\bf 718} (2013) 795
  [arXiv:1210.5482 [nucl-ex]].

\bibitem{Abelev:2012ola}
  B.~Abelev {\it et al.}  [ALICE Collaboration],
  Phys.\ Lett.\ B {\bf 719}  (2013) 29
  [arXiv:1212.2001 [nucl-ex]].

\bibitem{Aad:2012gla}
  G.~Aad {\it et al.}  [ATLAS Collaboration],
  Phys.\ Rev.\ Lett.\  {\bf 110}  (2013) 182302
  [arXiv:1212.5198 [hep-ex]].

\bibitem{Aaij:2015qcq} 
  R.~Aaij {\it et al.} [LHCb Collaboration],
  Phys.\ Lett.\ B {\bf 762}  (2016) 473
  [arXiv:1512.00439 [nucl-ex]].
  
\bibitem{Khachatryan:2016ibd}
  V.~Khachatryan {\it et al.} [CMS Collaboration],
  Phys.\ Rev.\ C {\bf 96} (2017) no.1,  014915
  [arXiv:1604.05347 [nucl-ex]].

\bibitem{Adare:2014keg}
  A.~Adare {\it et al.} [PHENIX Collaboration],
  Phys.\ Rev.\ Lett.\  {\bf 114} (2015) no.19,  192301
  [arXiv:1404.7461 [nucl-ex]].
  
\bibitem{Adamczyk:2015xjc}
  L.~Adamczyk {\it et al.} [STAR Collaboration],
  Phys.\ Lett.\ B {\bf 747} (2015) 265
  [arXiv:1502.07652 [nucl-ex]].
  
\bibitem{Adare:2015ctn}
  A.~Adare {\it et al.} [PHENIX Collaboration],
  Phys.\ Rev.\ Lett.\  {\bf 115} (2015) no.14,  142301
  [arXiv:1507.06273 [nucl-ex]].

\bibitem{Khachatryan:2016txc}
  V.~Khachatryan {\it et al.} [CMS Collaboration],
  Phys.\ Lett.\ B {\bf 765} (2017) 193
  [arXiv:1606.06198 [nucl-ex]].
    
\bibitem{Aaboud:2016yar}
  M.~Aaboud {\it et al.} [ATLAS Collaboration],
  Phys.\ Rev.\ C {\bf 96} (2017) no.2,  024908
  [arXiv:1609.06213 [nucl-ex]].
  
\bibitem{Aaboud:2017acw}
  M.~Aaboud {\it et al.} [ATLAS Collaboration],
  Eur.\ Phys.\ J.\ C {\bf 77} (2017) no.6,  428
  [arXiv:1705.04176 [hep-ex]].
  
\bibitem{Aaboud:2017blb}
  M.~Aaboud {\it et al.} [ATLAS Collaboration],
  Phys.\ Rev.\ C {\bf 97} (2018) no.2,  024904
  [arXiv:1708.03559 [hep-ex]].
  
\bibitem{Bozek:2012gr}
  P.~Bozek and W.~Broniowski,
  Phys.\ Lett.\ B {\bf 718}  (2013) 1557
  [arXiv:1211.0845 [nucl-th]].
  
\bibitem{Shuryak:2013ke}
  E.~Shuryak and I.~Zahed,
  Phys.\ Rev.\ C {\bf 88} (2013) 044915
  [arXiv:1301.4470 [hep-ph]].
  
\bibitem{Bzdak:2013zma}
  A.~Bzdak, B.~Schenke, P.~Tribedy and R.~Venugopalan,
  Phys.\ Rev.\ C {\bf 87}  (2013) 064906
  [arXiv:1304.3403 [nucl-th]].
  
\bibitem{Werner:2010ss}
  K.~Werner, I.~Karpenko and T.~Pierog,
  Phys.\ Rev.\ Lett.\  {\bf 106} (2011) 122004
  [arXiv:1011.0375 [hep-ph]].

\bibitem{Gavin:2008ev}
  S.~Gavin, L.~McLerran and G.~Moschelli,
  Phys.\ Rev.\ C {\bf 79}  (2009) 051902
  [arXiv:0806.4718 [nucl-th]].
  
\bibitem{Chesler:2009cy}
  P.~M.~Chesler and L.~G.~Yaffe,
  Phys.\ Rev.\ D {\bf 82} (2010) 026006
  [arXiv:0906.4426 [hep-th]].
  
\bibitem{Heller:2011ju}
  M.~P.~Heller, R.~A.~Janik and P.~Witaszczyk,
  Phys.\ Rev.\ Lett.\  {\bf 108} (2012) 201602
  [arXiv:1103.3452 [hep-th]].
  
\bibitem{Kurkela:2015qoa}
  A.~Kurkela and Y.~Zhu,
  Phys.\ Rev.\ Lett.\  {\bf 115} (2015) no.18,  182301
  [arXiv:1506.06647 [hep-ph]].
  
\bibitem{Romatschke:2016hle}
  P.~Romatschke,
  Eur.\ Phys.\ J.\ C {\bf 77} (2017) no.1,  21
  [arXiv:1609.02820 [nucl-th]].
  
\bibitem{Gelis:2010nm}
  F.~Gelis, E.~Iancu, J.~Jalilian-Marian and R.~Venugopalan,
  Ann.\ Rev.\ Nucl.\ Part.\ Sci.\  {\bf 60} (2010) 463
  [arXiv:1002.0333 [hep-ph]].
  
\bibitem{Hwa:2008um}
  C.~B.~Chiu, R.~C.~Hwa and C.~B.~Yang,
  Phys.\ Rev.\ C {\bf 78}  (2008) 044903
  [arXiv:0801.2183 [nucl-th]].

\bibitem{Bjorken:2013boa}
  J.~D.~Bjorken, S.~J.~Brodsky and A.~Scharff Goldhaber,
  Phys.\ Lett.\ B {\bf 726}  (2013) 344
  [arXiv:1308.1435 [hep-ph]].

\bibitem{Shuryak:2013sra}
  E.~Shuryak and I.~Zahed,
  Phys.\ Rev.\ D {\bf 89} (2014) 094001
  [arXiv:1311.0836 [hep-ph]].

\bibitem{Andres:2014bia}
  C.~Andres, A.~Moscoso and C.~Pajares,
  Phys.\ Rev.\ C {\bf 90} (2014) 054902
  [arXiv:1405.3632 [hep-ph]].

\bibitem{Armesto:2006bv}
  N.~Armesto, L.~McLerran and C.~Pajares,
  Nucl.\ Phys.\ A {\bf 781} (2007) 201
  [hep-ph/0607345].

\bibitem{Dumitru:2008wn}
  A.~Dumitru, F.~Gelis, L.~McLerran and R.~Venugopalan,
  Nucl.\ Phys.\ A {\bf 810} (2008) 91
  [arXiv:0804.3858 [hep-ph]].
  
\bibitem{Dumitru:2010iy}
  A.~Dumitru, K.~Dusling, F.~Gelis, J.~Jalilian-Marian, T.~Lappi and R.~Venugopalan,
  Phys.\ Lett.\ B {\bf 697} (2011) 21
  [arXiv:1009.5295 [hep-ph]].
  
\bibitem{Kovchegov:2012nd}
  Y.~V.~Kovchegov and D.~E.~Wertepny,
  Nucl.\ Phys.\ A {\bf 906} (2013) 50
  [arXiv:1212.1195 [hep-ph]].
  
\bibitem{Kovchegov:2013ewa}
  Y.~V.~Kovchegov and D.~E.~Wertepny,
  Nucl.\ Phys.\ A {\bf 925} (2014) 254
  [arXiv:1310.6701 [hep-ph]].
  
\bibitem{Dusling:2012iga}
  K.~Dusling and R.~Venugopalan,
  Phys.\ Rev.\ Lett.\  {\bf 108} (2012) 262001
  [arXiv:1201.2658 [hep-ph]].
  
\bibitem{Dusling:2012cg}
  K.~Dusling and R.~Venugopalan,
  Phys.\ Rev.\ D {\bf 87} (2013) no.5,  051502
  [arXiv:1210.3890 [hep-ph]].
  
\bibitem{Dusling:2012wy}
  K.~Dusling and R.~Venugopalan,
  Phys.\ Rev.\ D {\bf 87} (2013) no.5,  054014
  [arXiv:1211.3701 [hep-ph]].
  
\bibitem{Dusling:2013qoz}
  K.~Dusling and R.~Venugopalan,
  Phys.\ Rev.\ D {\bf 87} (2013) no.9,  094034
  [arXiv:1302.7018 [hep-ph]].
  
\bibitem{Dusling:2017dqg}
  K.~Dusling, M.~Mace and R.~Venugopalan,
  Phys.\ Rev.\ Lett.\  {\bf 120} (2018) no.4,  042002
  [arXiv:1705.00745 [hep-ph]].
  
\bibitem{Dusling:2017aot}
  K.~Dusling, M.~Mace and R.~Venugopalan,
  Phys.\ Rev.\ D {\bf 97} (2018) no.1,  016014
  [arXiv:1706.06260 [hep-ph]].
  
\bibitem{Blok:2017pui}
  B.~Blok, C.~D.~J\"akel, M.~Strikman and U.~A.~Wiedemann,
  JHEP {\bf 1712} (2017) 074
  [arXiv:1708.08241 [hep-ph]].
    
\bibitem{Skokov:2014tka}
  V.~Skokov,
  Phys.\ Rev.\ D {\bf 91} (2015) no.5,  054014
  [arXiv:1412.5191 [hep-ph]].
  
\bibitem{Schenke:2015aqa}
  B.~Schenke, S.~Schlichting and R.~Venugopalan,
  Phys.\ Lett.\ B {\bf 747} (2015) 76
  [arXiv:1502.01331 [hep-ph]].
    
\bibitem{Lappi:2015vta}
  T.~Lappi, B.~Schenke, S.~Schlichting and R.~Venugopalan,
  JHEP {\bf 1601} (2016) 061
  [arXiv:1509.03499 [hep-ph]].
  
\bibitem{McLerran:2016snu}
  L.~McLerran and V.~Skokov,
  Nucl.\ Phys.\ A {\bf 959} (2017) 83
  [arXiv:1611.09870 [hep-ph]].
  
\bibitem{Kovner:2016jfp}
  A.~Kovner, M.~Lublinsky and V.~Skokov,
  Phys.\ Rev.\ D {\bf 96} (2017) no.1,  016010
  [arXiv:1612.07790 [hep-ph]].
  
\bibitem{Kovchegov:2018jun}
  Y.~V.~Kovchegov and V.~V.~Skokov,
  arXiv:1802.08166 [hep-ph].
  
\bibitem{Altinoluk:2016vax}
  T.~Altinoluk, N.~Armesto, G.~Beuf, A.~Kovner and M.~Lublinsky,
  Phys.\ Rev.\ D {\bf 95} (2017) no.3,  034025
  [arXiv:1610.03020 [hep-ph]].
  
\bibitem{Kovner:2017gab}
  A.~Kovner, M.~Lublinsky and V.~Skokov,
  Phys.\ Rev.\ D {\bf 96} (2017) no.9,  096003
  [arXiv:1706.02330 [hep-ph]].
    
\bibitem{Martinez:2018ygo}
  M.~Martinez, M.~D.~Sievert and D.~E.~Wertepny,
  arXiv:1801.08986 [hep-ph].

\bibitem{Kovner:2017vro}
  A.~Kovner and A.~H.~Rezaeian,
  Phys.\ Rev.\ D {\bf 95} (2017) no.11,  114028
  [arXiv:1701.00494 [hep-ph]].

\bibitem{Kovner:2017ssr}
  A.~Kovner and A.~H.~Rezaeian,
  Phys.\ Rev.\ D {\bf 96} (2017) no.7,  074018
  [arXiv:1707.06985 [hep-ph]].
  
\bibitem{Kovner:2018vec}
  A.~Kovner and A.~H.~Rezaeian,
  arXiv:1801.04875 [hep-ph].
  
\bibitem{Kovner:2010xk}
  A.~Kovner and M.~Lublinsky,
  Phys.\ Rev.\ D {\bf 83} (2011) 034017
  [arXiv:1012.3398 [hep-ph]].
  
\bibitem{Kovner:2011pe}
  A.~Kovner and M.~Lublinsky,
  Phys.\ Rev.\ D {\bf 84} (2011) 094011
  [arXiv:1109.0347 [hep-ph]].
  
\bibitem{Kovner:2012jm}
  A.~Kovner and M.~Lublinsky,
  Int.\ J.\ Mod.\ Phys.\ E {\bf 22} (2013) 1330001
  [arXiv:1211.1928 [hep-ph]].
  
\bibitem{Dumitru:2014vka}
  A.~Dumitru and V.~Skokov,
  Phys.\ Rev.\ D {\bf 91} (2015) no.7,  074006
  [arXiv:1411.6630 [hep-ph]].
  
\bibitem{Levin:2011fb}
  E.~Levin and A.~H.~Rezaeian,
  Phys.\ Rev.\ D {\bf 84} (2011) 034031
  [arXiv:1105.3275 [hep-ph]].
  
\bibitem{Altinoluk:2015uaa}
  T.~Altinoluk, N.~Armesto, G.~Beuf, A.~Kovner and M.~Lublinsky,
  Phys.\ Lett.\ B {\bf 751} (2015) 448
  [arXiv:1503.07126 [hep-ph]].
  
\bibitem{Altinoluk:2015eka}
  T.~Altinoluk, N.~Armesto, G.~Beuf, A.~Kovner and M.~Lublinsky,
  Phys.\ Lett.\ B {\bf 752} (2016) 113
  [arXiv:1509.03223 [hep-ph]].


\bibitem{GolecBiernat:1998js} 
  K.~J.~Golec-Biernat and M.~Wusthoff,
  Phys.\ Rev.\ D {\bf 59}  (1998) 014017
  [hep-ph/9807513].




\end{thebibliography}
\end{document}